\documentclass{pasa}%

\pdfoutput=1

\title{A Framework for HI Spectral Source Finding Using Distributed-Memory Supercomputing}
\author[Stefan Westerlund and Christopher Harris]{Stefan Westerlund$^{1,3}$ \and Christopher Harris$^2$\\
\affil{$^1$ICRAR/University of Western Australia, M468 35 Stirling Highway, Crawley, WA 6009}%
\affil{$^2$iVEC@UWA/University of Western Australia, M024 35 Stirling Highway, Crawley, WA 6009}
\affil{$^3$Email: stefan.westerlund@icrar.org}}%
\jid{PASA}
\doi{10.1017/pas.\the\year.xxx}
\jyear{\the\year}

\usepackage[authoryear]{natbib}
\bibpunct{(}{)}{;}{a}{}{,}
\usepackage{graphicx}
\usepackage[small,labelfont=bf]{caption}
\usepackage{subcaption}
\usepackage[obeyspaces]{url}
\usepackage{aas_macros}

\begin{document}%
\begin{abstract}
The latest generation of radio astronomy interferometers will conduct all sky surveys with data products consisting of petabytes of spectral line data. Traditional approaches to identifying and parameterising the astrophysical sources within this data will not scale to datasets of this magnitude, since the performance of workstations will not keep up with the real-time generation of data. For this reason, it is necessary to employ high performance computing systems consisting of a large number of processors connected by a high-bandwidth network. In order to make use of such supercomputers substantial modifications must be made to serial source finding code. To ease the transition, this work presents the Scalable Source Finder Framework, a framework providing storage access, networking communication and data composition functionality, which can support a wide range of source finding algorithms provided they can be applied to subsets of the entire image. Additionally, the Parallel Gaussian Source Finder was implemented using SSoFF, utilising Gaussian filters, thresholding, and local statistics. PGSF was able to search on a 256GB simulated dataset in under $24$ minutes, significantly less than the 8 to 12 hour observation that would generate such a dataset.
\end{abstract}
\maketitle%

\section{INTRODUCTION}
A critical stage of radio astronomy spectral-line image analysis is source finding, which identifies the galaxies present in the image and determines their position and other parameters. As surveys increase in size, with larger fields of view and greater resolution, they produce greater amounts of data. For example, the HIPASS survey \citep{HIPASSCatalogue} produced a total of 22GB of image data. By comparison the Widefield ASKAP L-band Legacy All-sky  Blind surveY (WALLABY) survey using the Australian Square Kilometre Array Pathfinder (ASKAP) telescope is expected to produce files of at least 256GB every 8 to 12 hours, with the entire all sky survey likely to total several petabytes.

%

Using a traditional desktop computer to perform source finding for these larger surveys is not feasible due to a number of factors, including processing rate, memory footprint and storage bandwidth. Extrapolating from test results, processing a $256$GB image using a single computer could take over $110$ hours to process on a single machine, if it could store the entire dataset in memory. The primary issue is that the numerical performance is not fast enough to keep up with the real-time data production of the telescope imaging pipeline. In addition to meeting real-time performance, the rate of source finding would ideally be significantly faster than the rate of production. This would allow reprocessing of the entire dataset should the source finder be improved during the survey.

Memory issues can also slow a source finder. If the machine running the source finder has insufficient physical RAM to store the data needed by the source finder, either the excess data will be stored on the hard disk, making access much slower, or the system will fail to allocate sufficient memory, halting the program. It is possible to write a source finder that only examines a portion of the image at a time, reducing the memory required, but this involves processing part or all of the image more than once. Because supercomputers have large amounts of memory available, it is more efficient to process the whole image at once.

Bandwidth to data storage may also limit performance, particularly if there is insufficient memory to hold the entire image. A single consumer hard disk can reach read data rates on the order of 100MB/s. To achieve higher bandwidths it will be necessary to use multiple disks, such as a RAID array or a parallel file system to have enough bandwidth available to read in an image sufficiently quickly.

In order to overcome these limitations, it is desirable to use multiple machines working together on the problem. Modern supercomputers consist of a cluster of computing nodes, where each node consists of one or more multi-core CPUs. A fast network is employed to connect the nodes to each other, and to a parallel file storage system. The scalability of the program across these nodes is important because future surveys will produce even greater amounts of data. It is desirable for the program to be able to expand and make effective use of a greater number of processors in order to search greater amounts of data.

However, in order for a source finding program to make use of such systems they must be written such that the data and processing are partitioned across the nodes, with communication via the network, and using parallel file operations. Additionally, specialised code libraries and application programming interfaces must be used such as MPI \citep{mpi}, MPI-IO and OpenMP \citep{openmp08}. Converting a serial program to run in parallel can thus take a significant amount of effort.

This work describes the Scalable Source Finding Framework (SSoFF), a framework with functionality to ease this transition. SSoFF handles the distribution of processing by dividing the image into portions and assigning them to a three-dimensional grid of processes. Each process performs the work required to search its portion of the image. SSoFF provides routines that allow the processes to read and write their portions of the image from the storage system, and to exchange intermediary values with their neighbouring processes.

With the functionality described above in place, existing source finding analysis routines can be adapted to process a portion of image data, and added to SSoFF. To demonstrate this, the Parallel Gaussian Source Finder (PGSF) was built using the framework. The analysis step of PGSF applies a series of three-dimensional, Gaussian filters to the data. For each filter, a threshold is applied based on the local data around each voxel, and voxels are selected if they are above the threshold for a set number of different filters. Additionally, voxel weightings can optionally be used if available.

Section 2 provides a background to source finding in radio astronomy. Section 3 presents each component of the framework in detail. Section 4 then details the implementation of PGSF. Section 5 provides benchmarking and correctness testing result, which are then discussed in Section 6. Finally, concluding remarks are included in Section 7.

\section{Background}
A source finder can form part of a pipeline for reducing and analysing data from telescopes. While the details of the configuration of the pipeline are highly specific to the instrument, survey and science goals, a general overview of the main stages for a interferometric, spectral-line HI survey are as follows. The first step is correlation, where the data from the different receivers are combined into visibilities. The imaging takes the calibrated visibilities and converts them into an image through a Fourier transformation. The imaging step also includes continuum subtraction, where unwanted continuum emission is removed from the data, and deconvolution, where sidelobes are removed from the image. Usually the source finder sits at the end of this pipeline, taking the images and searching them for sources, but in some cases they can be used to search the visibilities. The objects that are found are then measured to determine their properties, a process called parameterisation. The parameterised sources found by the source finder are then analysed to achieve the desired science goals for the survey.

The main measure of merit of a source finder is its accuracy, which has two components, completeness and reliability. \emph{Completeness} is the fraction of the sources in the image that have been found by the source finder. \emph{Reliability} is the fraction of sources reported by the source finder that are real sources in the image. Independent of the source finding accuracy is the accuracy of the parameterisation step.

The source finding framework presented in this work is intended for spectroscopic images of neutral hydrogen (HI) emission. A radio astronomy image, also known as a \emph{data cube}, breaks up the area of the sky being observed along three dimensions.  The first two are \emph{spatial dimensions} that denote the direction in the sky relating to a particular part of the image. The third dimension is frequency, which gives the frequency range of the observed radiation for a particular element of the image. For nearby sources this value may also be specified in terms of velocity, as frequency and velocity are related through the Doppler effect from the radial velocity of an object.

\subsection{Source Finding}
\begin{figure*}
	\centering
	\includegraphics[width=\textwidth]{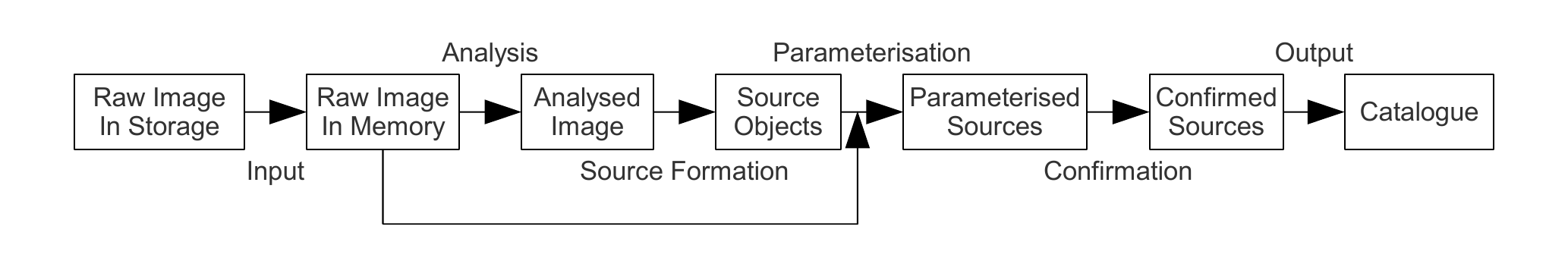}
	\caption{{\bf Source Finding Stages.} The radio image is read by the source finder and analysed to create a measure that states the likelihood that each voxel is part of a real source. The results of this analysis are used in the source formation step to select the set of voxels that are likely to be true sources, which then merges the chosen voxels together to form objects. The positions of these objects and the original image data are used to determine the parameters of the objects, and a confirmation step is applied to remove objects that appear to be false detections. The remaining objects and their parameters are produced as the output to the program.}
	\label{source_finder_flowchart}
\end{figure*}

The general process of searching an image is shown in \figurename~\ref{source_finder_flowchart}. The first step of the program is to read the image from storage into memory, in the \emph{input} step. This also involves any conversion of data to a format that the source finder uses.

The \emph{analysis} step applies a filter to the image, employs an analysis algorithm, or some combination of the two. The distinction used is that filtering techniques are algorithms that are designed to enhance signals based on their characteristics above the noise, whereas analysis techniques use statistical techniques to calculate the likelihood that a particular voxel is part of a real source. This step performs the bulk of the work and has the greatest diversity among the current serial source finding programs. \url{MultiFind} \citep{HIPASSCatalogue} searches for voxels that are above a threshold after Hanning smoothing the data and \url{Tophat} \citep{HIPASSCatalogue} searches for voxels that are above a threshold after convolving the data with top hat filters of different channel widths. \url{Duchamp} \citep{duchamp_paper} uses a choice of smoothing or the \emph{\`a trous} wavelet transform to reduce noise. The \url{2D-1D Wavelet} source finder \citep{2d_1d_wavelet_source_finder} uses 2D-1D wavelet transform to reduce noise, with the 2D transform operating in the spatial dimensions and the 1D transform operating in the spectral dimension. The \url{Smooth Plus Cut (S+C)} source finder \citep{s+c_source_finder} applies a series of different series of box filters, and takes the union of voxels that are above the threshold for each filter. The \url{Characterised Noise HI (CNHI)} source finder \citep{cnhi_source_finder} uses the Kuiper test to compare a test region of voxels to the noise of the image to locate regions with different flux properties, and the \url{Gamma Finder} \citep{WinkelThesis} uses the Gamma test to search for discontinuities in otherwise continuous, noisy data.


Voxels that are considered likely to be part of a true detection are selected from the results of the filtering or analysis, and then merged together to form objects in the \emph{source formation} step. For many filtering techniques, the selection of voxels involves calculating a threshold in flux or signal-to-noise ratio (SNR), and selecting voxels that have a value greater than this threshold. Some analysis techniques may effectively do this as part of their analysis algorithm.

The method used to decide whether or not to merge two voxels or groups of voxels can significantly affect the output of the source finder, particularly for sources that are only just above the detection limit. This merging can cause two types of errors, source confusion and source fragmentation. Source confusion occurs when two or more real objects are considered by the source finder to be the same object. Although HI sources are separated in three dimensions it is still possible for confusion to occur, depending on the proximity of the objects and the resolution of the image. Source fragmentation occurs when a source finder splits up a real object into two or more objects.

The objects that have been created in the source formation step are measured, using the original image data, to determine their parameters in the \emph{parameterisation} step. These parameters include the position, size and brightness of the objects, and will be used to study the galaxies, and other objects, found in the image. Parameterisation can be considered a separate, though related, problem to source finding, where source finding involves locating the sources of emission in an image, and parameterisation measures the properties of the sources. Another reason that parameterisation may be considered as separate to source finding is that once the source object positions are known, several different parameterisation techniques may be employed.

With the parameters of the objects known, the objects are passed through a \emph{confirmation} step. This component of the program examines the properties of each object and removes those whose properties suggest that they are likely to be false detections. The objects that survive the confirmation step are finally written out in the \emph{output} step of the program.

There is limited work currently published on applying source finding using HPC techniques. \citet{askap_source_finding} describes \url{Selavy}, a parallelisation of the \url{Duchamp} source finder. The framework presented in the following section is intended to make such parallel source finders easier to implement.

\section{FUNCTIONALITY OF SSoFF}
The Scalable Source Finding Framework (SSoFF) assists the development of parallel HI spectral-line source finders for High Performance Computing (HPC) systems by providing a number of components. These include work distribution, file IO, inter-process communication, statistics functions, voxel merging, and program control. Each of these components are described in greater detail in this section.

\subsection{Work Distribution}
The basis of a parallel program is organising multiple processes to work together and share a computational load. This component of SSoFF arranges the processes into a three dimensional grid, with $N_{x}$, $N_{y}$ and $N_{z}$ processes in each dimension. The image data is divided along the same three axes, right ascension, declination, and either frequency or velocity, into a number of portions equal to the number of processes along that side. That is, if the image has a total of $D_{x}$, $D_{y}$, and $D_{z}$ voxels along each size, then each process has $d_{t}$ voxels to process with $d_{x}$, $d_{y}$, and $d_{z}$ voxels along each side, according to the equations:

\begin{eqnarray}
	d_{t} &=& d_{x} d_{y} d_{z} \\
	d_{x} &=& \frac{D_{x}}{N_{x}} \\
	d_{y} &=& \frac{D_{y}}{N_{y}} \\
	d_{z} &=& \frac{D_{z}}{N_{z}}
\end{eqnarray}

The amount of voxels per side may vary by one between processes if the number of processes in the grid does not evenly divide the number of voxels in the image. Each portion of the image is assigned to its corresponding process to be searched. This arrangement allows for analysis algorithms that evaluate a voxel based on the properties of its surrounding voxels. The particular values of $N_{x}$, $N_{y}$, and $N_{z}$ may be set by the user of the program. The optimal values for performance may vary depending on the analysis techniques used, but it is often optimal to arrange them such that it minimises the amount of extra data that the system needs to store.

\begin{figure}
	\centering
	\includegraphics[width=\columnwidth]{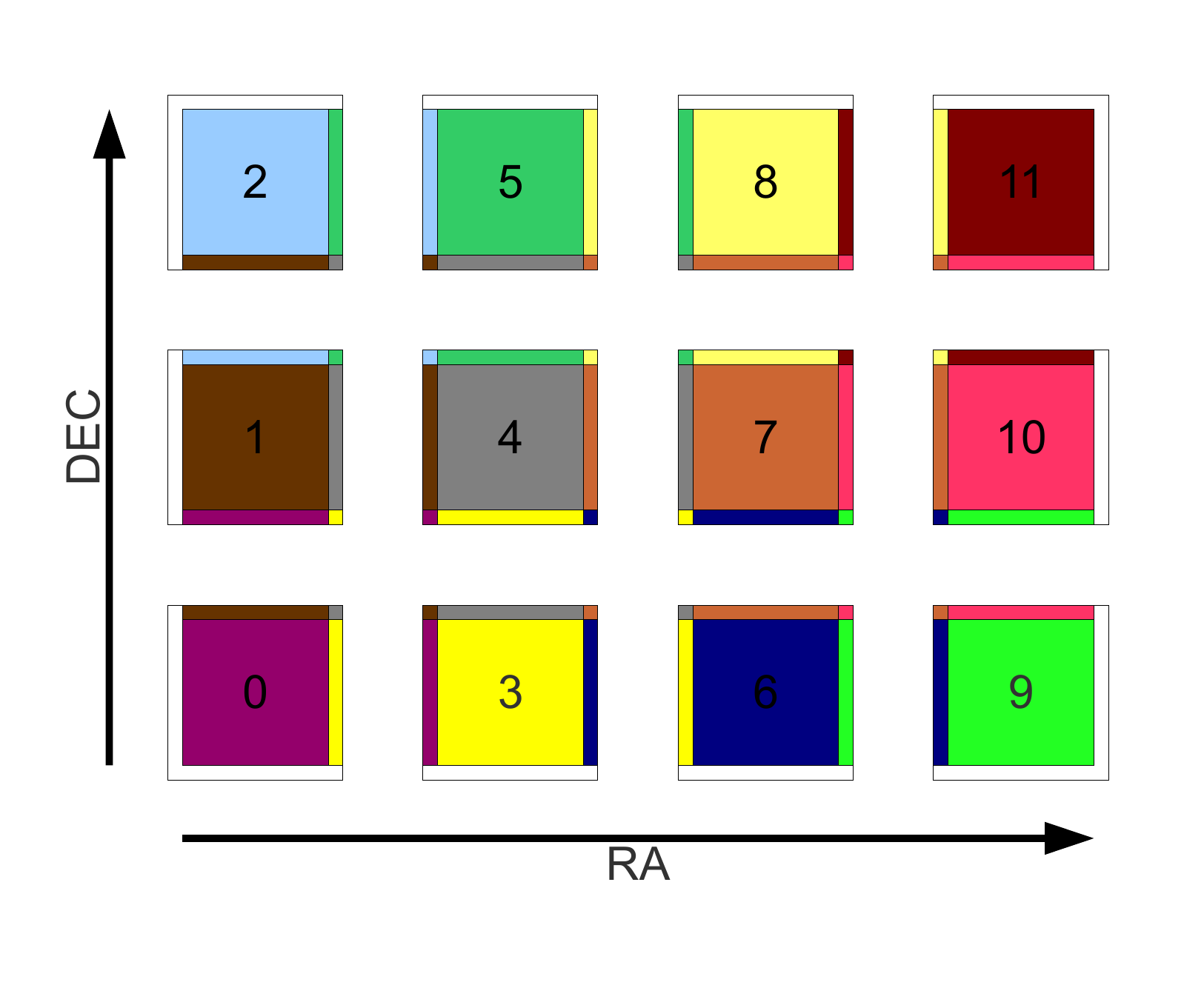}
	\caption{{\bf Data Distribution between Nodes.} Each node has its own section of the data cube, along with a portion from each adjacent node, so that it can correctly process the data assigned to it. The colour of the data indicates which node that data was assigned to. The white sections are padding that is placed around the edge of the image, so that the edge nodes can be processed in the same manner as the interior nodes. The division and allocation of image data to the processes can be changed, which may affect the computational performance of a source finder, but not its accuracy. The data can also be partitioned along the spectral axis, which is not shown here for clarity.}
	\label{data_distribution}
\end{figure}

When an analysis algorithm needs data from the neighbours of a particular voxel, this means that processing the voxels near the edge of an image portion will require image data that has been assigned to a different process. This information is provided by duplicating the voxel data from the edge of one process to another, such that each process holds a copy of the image data that is within a certain radius of its assigned portion. This extra information is known as \emph{halo data}. Algorithms that use the halo data will require the size of the halo to have a certain minimum size. If multiple algorithms use the halo data, the halo must be large enough for each of those algorithms. SSoFF provides data structures for each process to store its assigned image data and halo data as a three-dimensional array. This array data structure is used for the initial image data, intermediate values and the processed image. The manner in which the image data is divided between processes and stored is shown in \figurename~\ref{data_distribution}.

\subsection{File IO}
SSoFF provides methods to transfer data between a storage device, typically a hard drive, and the memory of the program. These functions will read the data in from the storage system and give each process its assigned portion of the image. Currently, the framework supports reading from a flat binary file, with three integers stating the size of the file along each axis, followed by the specified number of single-precision floating point numbers in row-major order. Support for additional file formats can be easily added as needed to SSoFF, because the file format is irrelevant to the framework once the data has been loaded into the image data structure.

The image data structure and the file IO methods are also capable of reading large files. Files greater than four gigabytes in size are too large to have each byte addressed using a 32 bit integer, so programs using those to address a file may be unable to properly read in and access the entire file. This function also provides a convenient way to bypass a limitation in MPI-IO, where it can only read in ~2GB of data per call. The data input function used by SSoFF avoids this problem by using multiple MPI-IO function calls to read the data. The data structures used by SSoFF to store the image data are addressed using three 32-bit integers, so the entire image can be accessed using integers as long as each side length is less than $2^{31} - 1$ elements, the maximum value of a signed integer. Alternatively, functions that use the data structures provided by SSoFF may choose to access them using a single 64-bit integer index.

The halo data for the edges of the cube is set to zero. When using weighted calculations, this will automatically assign those voxels a weight of zero. For unweighted calculations, these voxels are treated the same as the voxels from the image. The effect of the padded values is left to be considered in future work.

There are two options of ensuring that each node has the halo data it needs. Either the nodes read both their assigned data and the halo data directly from the storage system, or they read only their assigned data from storage, then use the network to transfer the halo data. Because the same network is used to transfer data from the storage system and inter-node communication, both of these methods would result in the same performance if the network is the bottleneck. However, if the storage system is the bottleneck in the transfer, then the second method will be faster because it reads less data from storage. Therefore, the framework uses the network to transfer the halo data, as covered in the next section.



\subsection{Inter-process Communication}
\begin{figure}
	\centering
	\begin{subfigure}{\columnwidth}
		\centering
		\includegraphics[width=0.75\columnwidth]{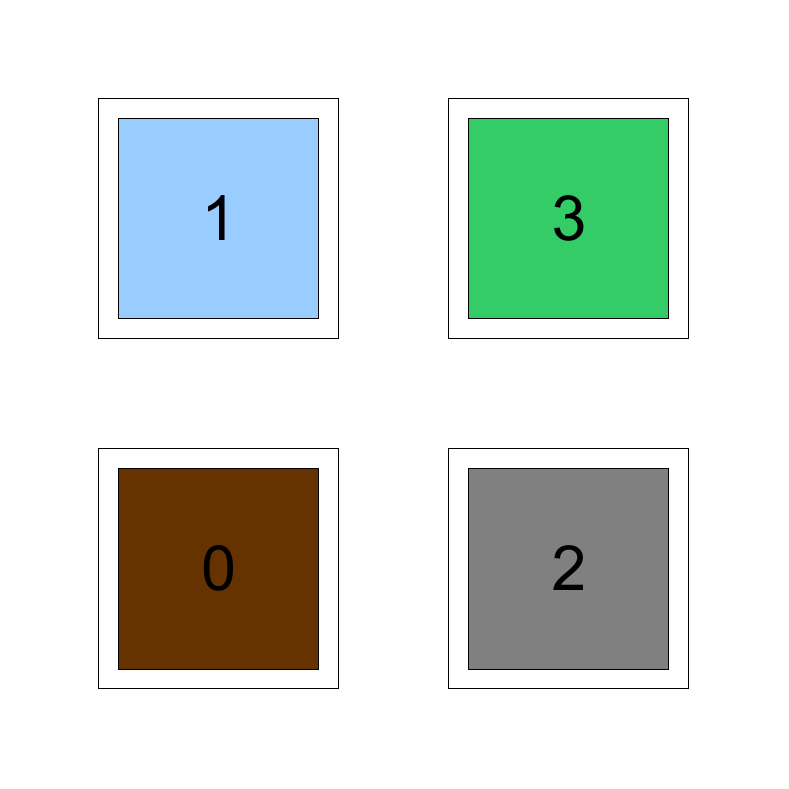}
		\caption{Initial State}
	\end{subfigure}
	
	\begin{subfigure}{\columnwidth}
		\centering
		\includegraphics[width=0.75\columnwidth]{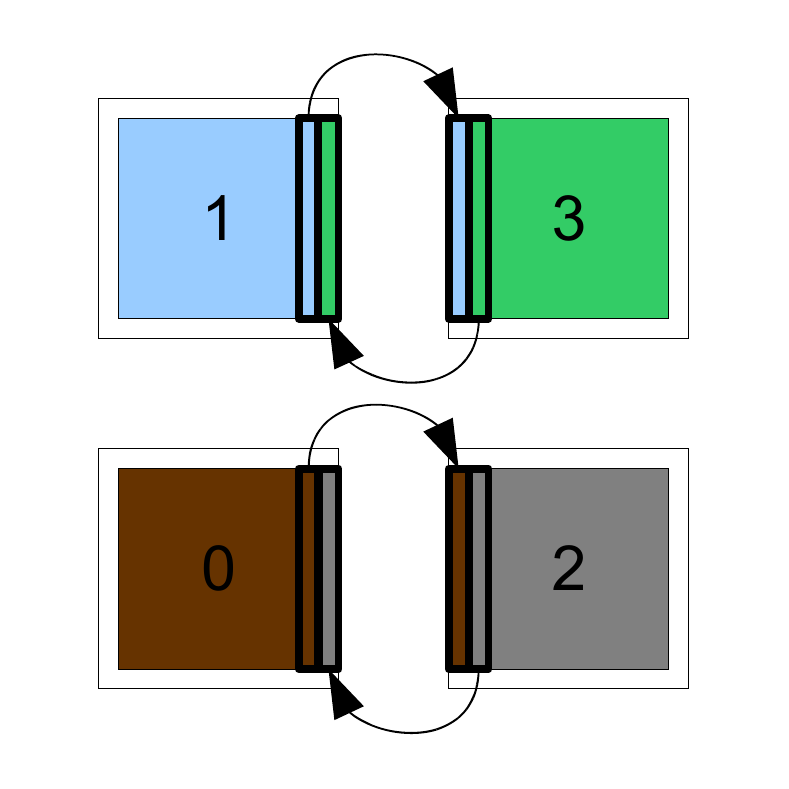}
		\caption{Transfer in x Direction}
	\end{subfigure}
	
	\begin{subfigure}{\columnwidth}
		\centering
		\includegraphics[width=0.75\columnwidth]{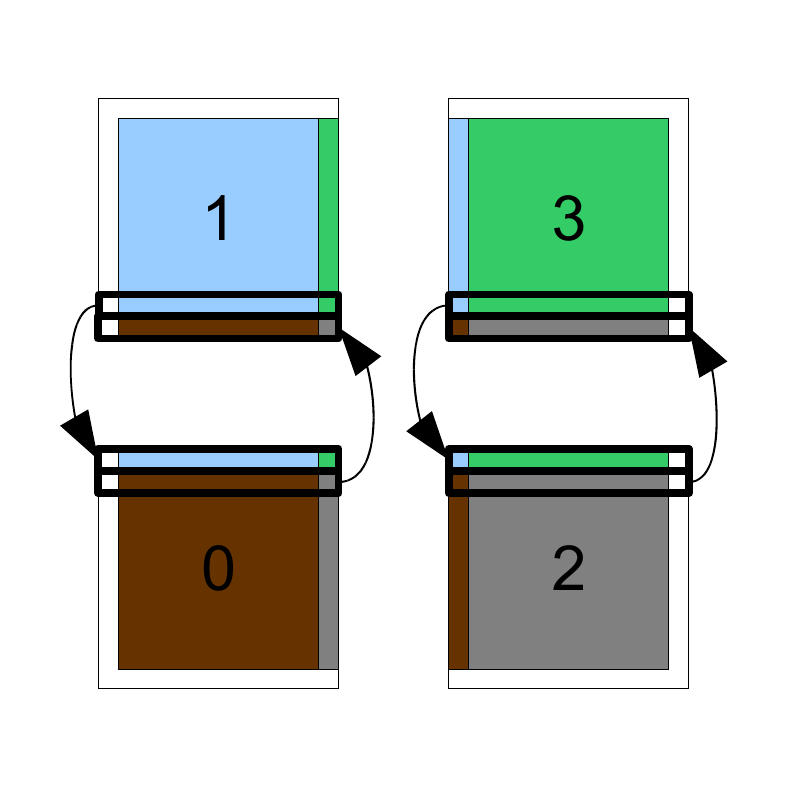}
		\caption{Transfer in y Direction}
	\end{subfigure}
	
	\caption{{\bf Halo Communication.} SSoFF transfers halo data in three steps, one for each axis. The bold lines show the data transferred in each step. The processes send the data they hold, as well as data that they received in previous steps. Only the x axis and y axis transfers are shown here, the framework also does a third transfer along the z axis.}
	\label{halo_communication}
	
\end{figure}

Image analysis algorithms may require the image values around a voxel, in order to evaluate that voxel. In order for a process to analyse voxels near its border, it will require information that was assigned to its neighbouring processes. SSoFF provides a function to copy data from one process to the appropriate position in the halo data of its surrounding processes, using the array data structures mentioned above. The transfer is performed in three steps, as shown in \figurename~\ref{halo_communication}. First the halo data in the x axis is transferred, to the processes' left and right neighbours. Once this is complete, data is transferred in the y axis, to the top and bottom neighbours, including sending data that was received in the x axis transfer. Finally, the data is transferred in the z axis, between the front and back neighbours. Transferring data that was received from other nodes, in addition to data from a process's own node, ensures that processes still get the data they need even when they are not adjacent in the process grid.

The amount of data that is transferred by this function is dependent on the size of the data, the size of the halo per node, and the dimensions of the process grid. If the size of the halo along each dimension is equal to $H_{x}$, $H_{y}$, and $H_{z}$, then the amount of data transfer that occurs in when exchanging halo data is $T_{x}$ elements in the first, x axis, transfer step, $T_{y}$ elements in the second step and $T_{z}$ elements in the third step for a total of $T_{t}$ elements transferred per node:

\begin{eqnarray}
	T_{t} &=& N_{x} + N_{y} + N_{z} \\
	T_{x} &=& H_{x} d_{y} d_{z} \\
	T_{y} &=& (d_{x} + H_{x}) H_{y} d_{z} \\
	T_{z} &=& (d_{x} + H_{x}) (d_{y} + H_{y}) H_{z}
	\label{transfer_amount}
\end{eqnarray}

\subsection{Statistics Functions}
Source finders often require statistical functions to provide a measure of the probability that a voxel is part of a valid source. SSoFF provides several statistics functions that operate across a distributed dataset. These include calculating the mean and standard deviation of the dataset, for both weighted and unweighted data. There is an option for global statistics, that calculate the values based on the entire contents of the array, and a local version that calculates the mean and standard deviation for each voxel individually based on the data within a user-specified range that is $L_x$, $L_y$ and $L_z$ voxels in size along the x, y, and z axes. Because the local statistics calculations use data from the surrounding nodes, these functions impose a minimum halo size equal to the range that the local statistics are being calculated across. The global statistics are calculated by each node determining the sum of its own portion of the data then using MPI to perform a global sum reduction across the different processes.

The calculation of the local statistics requires that the node possesses the values surrounding a voxel up to the specified range, so the halo size must be at least as large as the local statistics size, and the program must perform a halo transfer so that each process has the information it needs. Once this transfer is complete, each node can calculate the individual sum for each voxel. The mean, $\mu_w$, and the standard deviation, $\sigma_w$ for weighted data are calculated as shown in Equations~\ref{weighted_mean} and \ref{weighted_std_dev}, from the image data $d$ and the weights data $w$. In the case of unweighted data, the mean, $\mu_u$, and standard deviation, $\sigma_u$ are calculated as in Equations~\ref{unweighted_mean} and \ref{unweighted_std_dev}. With the mean and standard deviation values known, they are used to calculate the z score for each voxel in the data, as shown in Equations~\ref{weighted_z_score} and \ref{unweighted_z_score}.

\begin{align}
	\label{weighted_z_score}
	z_w[x][y][z] &= \frac{d[x][y][z] - \mu_w[x][y][z]}{\sigma_w[x][y][z]} \\
	\label{weighted_std_dev}
	\sigma_w[x][y][z] &= \sqrt{\frac{S_w[x][y][z] ~ W[x][y][z] - \mu_w[x][y][z]^{2}}{W[x][y][z]^{2}}} \\
	S_w[x][y][z] &= \sum\limits_{o_x} \sum\limits_{o_y} \sum\limits_{o_z} w[i_x][i_y][i_z] ~ d[i_x][i_y][i_z]^2 \\
	\label{weighted_mean}
	\mu_w[x][y][z] &= \sum\limits_{o_x} \sum\limits_{o_y} \sum\limits_{o_z} w[i_x][i_y][i_z] ~ d[i_x][i_y][i_z] \\
	W[x][y][z] &= \sum\limits_{o_x} \sum\limits_{o_y} \sum\limits_{o_z} w[i_x][i_y][i_z]
\end{align}

\begin{align}
	\label{unweighted_z_score}
	z_u[x][y][z] &= \frac{d[x][y][z] - \mu_u[x][y][z]}{\sigma_u[x][y][z]} \\
	\label{unweighted_std_dev}
	\sigma_u[x][y][z] &= \sqrt{\frac{S_u[x][y][z]}{C} - \mu_u[x][y][z]^{2}} \\
	S_u[x][y][z] &= \sum\limits_{o_x} \sum\limits_{o_y} \sum\limits_{o_z} d[i_x][i_y][i_z]^2 \\
	\label{unweighted_mean}
	\mu_u[x][y][z] &= \frac{\sum\limits_{o_x} \sum\limits_{o_y} \sum\limits_{o_z} d[i_x][i_y][i_z]}{C} \\
	C &= L_x L_y L_z \\
	i_x &= x + o_x, ~o_x \in [-\frac{L_x}{2}, \frac{L_x}{2}]\\
	i_y &= y + o_y, ~o_y \in [-\frac{L_y}{2}, \frac{L_y}{2}]\\
	i_z &= z + o_z, ~o_z \in [-\frac{L_z}{2}, \frac{L_z}{2}]
\end{align}

The totals are calculated using moving sums, tracking three values for $S_w$, $W$ and $\mu_w$. In order to minimise memory use the statistics are performed in-place, overwriting the original image values with the sigma value for each voxel. Small additional buffers are used to store intermediate results. Each of these temporary buffers are $(d_{x} + L_{x} - 1)(d_{y} + L_{y} - 1)\frac{L_{z}}{2}$ elements in size. This is the smallest possible size of the buffers because the algorithm overwrites the image data in the x direction as it progresses, so previous values along the x axis must be read from a buffer while following values can be read from the image array. Both the weighted and unweighted local statistics calculations use one buffer of floats to store the original image information, and two buffers of doubles to store the sums for the mean and standard deviation. The weighted calculation requires an additional buffer of doubles to store the summed weights.

The sums are first calculated as one-dimensional moving sums along the z axis. When initialising these sums, one multiply is required to calculate the value of $\mu_w$, two multiplies to calculate the squared value for $S_w$ and one addition each to update the three values. This results in a total of six floating point operations per voxel. Because the buffers are half the size of $L_z$ along the z axis each voxel is initialised twice, so this initialisation requires twelve floating point operations per voxel.

Once the moving sums have been initialised, it is executed across the buffer data, adding in new values and subtracting old values. This part of the algorithm requires twice as many calculations as the initialisation due to subtracting previous values but is only performed once per voxel. Therefore, performing the moving sum across the z axis requires a total of $24$ floating point operations per voxel. The z axis sums are performed on all of a node's assigned voxels plus a halo of $L_x$ voxels in the x axis and $L_y$ voxels in the y direction, because these values will be used in later sums. The number of z-axis operations required to calculate the local statistics could be reduced by using larger buffers, but this would come at the cost of increased memory use.

Once the z axis sums are complete, the sums are performed across the y axis, and then the x axis. These sums only require additions and subtractions, as all multiplications have been performed in calculating the sums across the z axis. Initialising these sums requires three operations per voxel, one addition for each of the three values. The y-axis initialisation only needs to be performed once for each x-z line in the image and is calculated over $L_y$ voxels per line. Likewise, the x-axis moving sum initialisation is performed once for each y-z line in the image, across $L_x$ voxels per line. Once initialised, performing the moving sums requires six floating point operations per voxel, as the previous values need to be subtracted from the sum. The y-axis sums are calculated across a node's assigned voxels, and an additional halo in the x axis. The x axis sums require no halo. With the sums calculated across the three axes, they are used to calculate the final z value. This requires five floating point operations, three divides and one square root per voxel assigned to a node.

The use of moving sums reduces the amount of computational effort required, but the statistics can still be a significant portion of a source finder's running time. The number of floating point operations required per node for the local mean and standard deviation for a weighted dataset is approximately equal to $O_{w, t}$:

\begin{eqnarray}
	\label{stats_op_count_weighted}
	O_{w, t} &\approx& O_{w, z} + O_{w, y} + O_{w, x} \\
	O_{w, z} &=& 24 (d_x + L_x - 1)(d_y + L_y - 1) d_z\\
	O_{w, y} &=& (d_x + L_x - 1) (3 L_y + 6 d_y) d_z \\
	O_{w, x} &=& (3 L_x + 6 d_x) d_y d_z
\end{eqnarray}

Calculating the local statistics in the unweighted case is similar to the weighted case, but fewer operations are required. Only two sums are tracked, $S_u$ and the numerator of $\mu_u$, and the weights values don't need to be multiplied into the sums. This halves the number of operations required for the z axis sums, and reduces the number of y and x axis operations by a third. Calculating the final z score requires three floating point operations, three divisions and one square root calculation per voxel. The total operation count per node to calculate the local statistics in the unweighted case is equal to $O_{u, t}$:

\begin{eqnarray}
	\label{stats_op_count_unweighted}
	O_{u, t} &\approx& O_{u, z} + O_{u, y} + O_{u, x} \\
	O_{u, z} &=& 12 (d_x + L_x - 1)(d_y + L_y - 1) d_z \\
	O_{u, y} &=& (d_x + L_x - 1) (2 L_y + 4 d_y - 1) d_z \\
	O_{u, x} &=& (2 L_x + 4 d_x) d_y d_z
\end{eqnarray}

\subsection{Source Formation}
A source finder must decide which voxels in an image are considered part of a legitimate source of emission and to collect these voxels into data objects that represent these sources. These are two separate but related tasks, called selection and merging. SSoFF provides functionality to perform these tasks in a parallel environment.

Selecting voxels is often done by applying a threshold to a dataset. This framework allows for thresholding across both the image array data structure, and across a sparse image dataset in the form of a hash map. These functions use a flood fill algorithm to pick the voxels that are above the threshold, and merge adjacent voxels into source objects.

The flood fill merges source objects within a single process but sources may be split across multiple processes. SSoFF merges these objects using a multi-step procedure. First each process sends the positions of voxels along its borders to its neighbouring processes. This is used to find where source objects are split between processes. Each part of a split object is given a destination process. The destination for a split source object is the process with a part of that object, that has the lowest index in the process grid. The destination process index is propagated across the different parts of a split source, from process to process, to ensure that each part of the source is sent to the same process.

\subsection{Program control}
There are a number of settings that can be changed to control how a source finder searches an image. These may include the choice of certain algorithms instead of others, and values to be used inside algorithms, in addition to specifying the data files to be used. SSoFF provides functionality to read in parameters from a file, using key-value pairs of strings to provide information. This also makes it easier to add new functions to a source finding program, as they can be added to the main routine and then check the contents of the parameter file to decide which functions to use at run time. The options can be used to specify values used inside the functions, such as the size of filters, or the value of the threshold to use when selecting voxels.

The framework makes use of several libraries to provide this functionality. \url{MPI} is used for the basis of the parallelisation, and to communicate data between processes. \url{SimCList} is used for linked lists, which are used to store collections of voxels and source objects, which vary in size depending on the contents of the dataset. \url{uthash} is used for hash tables, which are used to store sparse voxel information, and to store parameter file information. These libraries do not prevent source finders from using other libraries.

The functions described in this section provide a toolkit for writing a parallel source finder and can perform common source finding tasks in a parallel environment. Through this functionality, SSoFF reduces the difficulty of implementing additional functionality to a parallel source finder.  The use of this framework is demonstrated in the next section, where it is used to implement a source finder.

\section{Implementation of PGSF}
This section describes the Parallel Gaussian Source Finder (PGSF), a parallel source finder for HI spectral line images implemented using SSoFF. The analysis is based on the use of three-dimensional Gaussian filters, and voxels are selected if they are above the threshold for a set number of different filters. Sources constructed from these voxels are then subject to a confirmation step where only the sources whose spectral extent is is greater than the user-specified cut-off are written to the catalogue. PGSF can make use of an arbitrary number of processes, up to the number of voxels in the image being searched. This source finder can also process large files, limited by the memory of the nodes used to search the image and to a maximum size of $2^{31} - 1$ voxels in each dimension. It processes an image that consists of single-precision floating point numbers, but it can be easily extended to other data types. The details of PGSF are described below.

The {\bf analysis} algorithm used to inspect the image is a series of Gaussian filters probing different scales. It is based on the algorithm used by the \url{S+C} source finder \citep{s+c_source_finder} but has been expanded to run across parallel data, using SSoFF. A set of Gaussian filter templates are convolved with the data and the weights, as shown in Equation \ref{weighted_conv_eqn} for the weighted convolution where $F_x$, $F_y$ and $F_z$ are the dimensions of the filter template. If weights are unavailable an unweighted convolution is used, as shown in Equation \ref{conv_eqn}. As in the local statistics calculations, the output of the filter is only calculated for the voxels that have been assigned to a process, not for the process's halo values. As a result, filter output is only calculated once for values in the since the output for values in the halos are either calculated by the adjacent node that is responsible for that region, or not at all for the values outside the image. Additionally, the filtering process requires that the halo be at least the size of the largest filter used. When a process is filtering the edges of its assigned image data it uses the halo image values, and weights values if they are available, that were loaded in the input step of the program.


\begin{align}
	c_{w}[x][y][z] &= \notag\\
\label{weighted_conv_eqn}
	&\hspace{-0.16\columnwidth}\frac{\sum\limits_{p_x} \sum\limits_{p_y} \sum\limits_{p_z} d[j_x][j_y][j_z] ~ f[p_x][p_y][p_z] ~ w[j_x][j_y][j_z]}{\sum\limits_{p_x} \sum\limits_{p_y} \sum\limits_{p_z} w[j_x][j_y][j_z]} \\
\label{conv_eqn}
	c_u[x][y][z] &= \sum\limits_{p_x} \sum\limits_{p_y} \sum\limits_{p_z} d[j_x][j_y][j_z]~ f[p_x][p_y][p_z] \\
	j_x &= x + p_x, ~p_x \in [-\frac{F_x}{2},\frac{F_x}{2}] \\
	j_y &= y + p_y, ~p_y \in [-\frac{F_y}{2},\frac{F_y}{2}] \\
	j_z &= z + p_z, ~p_z \in [-\frac{F_z}{2},\frac{F_z}{2}]
\end{align}

For each filter the selection criteria used is a threshold equal to the mean plus a user-specified constant multiplied by the standard deviation, where the local mean and standard deviation are calculated from the filtered image, as in \citet{askap_source_finding}, using the local statistics functions of the framework. A count is kept for each voxel each time that voxel's filtered value above the threshold for a filter. After all of the filters have been applied, voxels are selected if their count is above a second user-defined threshold.

If a filter has a size of $F_x$, $F_y$ and $F_z$ elements, then the weighted version requires $4 D_x D_y D_z  F_x F_y F_z$ total floating point operations for that filter. This includes one multiply to combine the filter value and the weight for a voxel, one multiply to combine the filter weight value to the image value, one addition to update the sum of the convolution, and an addition to update the sum of the filter weight for that voxel. The unweighted version requires a total of $2 D_x D_y D_z  F_x F_y F_z$ operations, one multiply to combine the filter value and the image data value and one addition to update the sum of the filter values. This filtering does not require any data transfer between nodes beyond what is done reading in the image.

PGSF currently allows for arbitrary filter templates to be applied to the data. Ideally, a set of filter templates would be used that cover all possible sources, whilst limiting the amount of processing needed. Such an optimal set of filter templates has yet to be determined. Instead, a series of three-dimensional Gaussian functions are used. The sizes of these filters can be set by the user.

The {\bf parameterisation} of the sources is performed once they have been selected and merged by SSoFF. Because the data file format used by this program only stores the flux of the image, that is, the format stores no metadata, only a subset of the parameters can be determined. Performing a complete parameterisation of sources is considered to be outside the scope of this work. The parameters given by this program are the position of an object, as a flux-weighted mean in units of the array indices, the peak flux of the object and the sum of its flux across its voxels, and the width of the object along the spectral axis. PGSF also handles re-reading the information from the image cube when it is needed, to parameterise sources that contain voxels that were received from other nodes during the source formation step.

The {\bf confirmation} of sources makes use of the parameterisation information to confirm or reject potential sources. PGSF rejects sources that are below a user-specified channel width. This is because most legitimate sources have a spectral width that is significantly larger than the channel width of a spectral-line image, so sources that have a small channel width are likely to be noise peaks or interference. For example, the thinnest galaxy in the HIPASS Catalogue \cite{HIPASSCatalogue}, J1336-29 has a velocity width of $30.4$km/s \footnote{Using the measure $W_{50}^{max}$.} compared to the WALLABY survey, which will have a spectral resolution of $4$km/s \cite{WallabyProposal}. The framework allows for more complex confirmation techniques to be added. The confirmed sources and their parameters are written to the output catalogue.

PGSF can scale to search larger images, up to datasets that are $2^{31} - 1$ elements along each side. The number of processes to be used by the source finder has a upper limit equal to the number of voxels in the image, and a lower limit set by memory limits. Each process has a copy of its assigned portion of the image, including the halo data, which is $(d_x + H_x)(d_y + H_y)(d_z + H_z)$ elements per process. A second array data structure of the same size is used to store the filtered image, and optionally a third data structure to store the weights information of the image. Additional memory is used when calculating the statistics, as mentioned above, and a variable amount of memory is needed to store the source detections. The accuracy and computational performance of this program is measured in the next section.

\section{TESTING}
Several tests were employed to measure the correctness of PGSF. The first set of tests were performed to ensure that the program was working correctly. The second set of tests shows the accuracy of the source finder for different sources. Finally, the third set of tests analyse the suitability of the program for processing large datasets. The primary machine used to test the program was the Epic@Murdoch supercomputer. This system has $800$ nodes, each possessing two six-core Intel Xeon X5660 CPUS, 24GB of RAM and a QLogic IBA7322 QDR Infiniband interconnect. The MPI library used was OpenMPI version 1.6.3.

The correctness of the program was examined using unit testing. These tests do not concern the overall accuracy of the program. Rather, they show that the framework functions work correctly. Each function used in PGSF was individually tested for correctness. The correctness for the program as a whole was demonstrated by running the source finder on a 2GB simulated data cube and comparing the output to the expected results, which were obtained by executing a single-threaded implementation of the program on the same image. The results were found to be identical.

The accuracy of PGSF was demonstrated by executing the source finder on a pair of simulated data cubes, one containing point sources and one containing extended sources. These are the images used in \citet{source_finder_comparison}. \emph{Point sources} are objects whose spatial extent is smaller than the resolution of the instrument used to observe them, whilst \emph{extended sources} are those who are spatially larger than the resolution of the instrument. In practice, both types of sources have spectral sizes significantly larger than the channel width of the telescope. This test used a series of filters that are 1, 3, 5 and 9 pixels wide in the spatial dimensions and 9, 17, 33, 65 and 129 channels wide in the spectral dimension.  For the data cube used, these filters are 10", 30", 50" and 90" in spatial size, and 659kHz, 1.24MHz, 2.42MHz, 4.76MHz and 9.45MHz in spectral size. In all, 20 different filters were used with a total of $29,348$ filter elements between them. For each filter, a threshold equal to four standard deviations above the mean is applied to the filtered data. As in \citet{askap_source_finding}, the mean and standard deviation for a voxel are calculated from a range $101$ voxels wide in right ascension and declination, and a single voxel wide in frequency. Detected voxels were merged into the same object if they were within $5$ voxels spatially and $80$ frequency channels of another voxel in that object. The final selection of voxels were those that were above the threshold for $13$ or more filters, and had a spectral width of at least $10$ channels.

Both test images have corresponding mask files that show where the real sources exist in the image. These masks, combined with the original image data, were converted into catalogues using an external parameterisation script, as described in \citet{cnhi_source_finder}. These catalogues are used as the reference point for the sources that are in the images. The program was then run on the two images and produced its own mask file for each image. These mask files were also converted into catalogues by the parameterisation script. Using an external script allows for a more detailed parameterisation of the sources than that provided by the algorithms currently implemented in the program.

The detected catalogues were cross-matched against the reference catalogues using the Source Finder Accuracy Evaluator (SFAE) \citep{sfae}. The accuracy of the source finder can be seen in the results of the cross-matching, as shown in \figurename~\ref{accuracy}. Figures \ref{point_completeness} and \ref{point_reliability} show the accuracy of the source finder for point sources, and Figures \ref{extended_completeness} and \ref{extended_reliability} show the accuracy for extended sources. These are based on the point-source and extended-source simulated cubes, respectively. Figures~\ref{point_completeness} and \ref{extended_completeness} use the reference catalogue for the peak flux values and source counts, whilst Figures\ref{point_reliability} and \ref{extended_reliability} use the results of the source finder for their values.

\begin{figure*}
	\centering
	\begin{subfigure}{\columnwidth}
		\resizebox{\columnwidth}{!}{\input{./point_binned_completeness_by_snr_tex}}
		\caption{Point Source Completeness}
		\label{point_completeness}
	\end{subfigure}
	\begin{subfigure}{\columnwidth}
		\resizebox{\columnwidth}{!}{\input{./point_binned_reliability_by_snr_tex}}
		\caption{Point Source Reliability}
		\label{point_reliability}
	\end{subfigure}
	\\
	\begin{subfigure}{\columnwidth}
		\resizebox{\columnwidth}{!}{\input{./extended_binned_completeness_by_snr_tex}}
		\caption{Extended Source Completeness}
		\label{extended_completeness}
	\end{subfigure}
	\begin{subfigure}{\columnwidth}
		\resizebox{\columnwidth}{!}{\input{./extended_binned_reliability_by_snr_tex}}
		\caption{Extended Source Reliability}
		\label{extended_reliability}
	\end{subfigure}
	\caption{{\bf Source Finder Accuracy.} These plots show the completeness and reliability of PGSF for different sources. The abscissa is the peak SNR  of the bins, equal to the peak flux of a source, divided by the image's RMS value. The reference catalogue's value for the peak SNR is used for the completeness plots and the detected catalogue's value is used for the reliability plots. The histogram shows the completeness and reliability for the sources in that bin, with the error bars showing a one-sigma error calculated using bootstrap resampling. The dotted green line shows the number of sources in each bin. The point source results were obtained from an image that contained only point sources, likewise the extended source results were obtained using an image that contained only extended sources. These values were obtained using a threshold of four$\sigma$, and voxels that were above that threshold for $13$ different filters.}
	\label{accuracy}
\end{figure*}

PGSF is intended for use on images of the same size as those that will be used in the WALLABY survey. The size of these images will depend on the configuration of the telescope, but a likely data size is approximately $256$GB, consisting of $2048 \times 2048$ spatial values and $16,384$ frequency channels, with each value stored as a four-byte floating point number. This image will not have polarisation information, but it may have a weighting cube, for an additional $256$GB. Data from ASKAP is not yet available, so a placeholder image was created from a $64$GB simulated image\footnote{This image is available from \url{http://www.atnf.csiro.au/people/Matthew.Whiting/ASKAPsimulations.php}, Set~\#7, made by combining the ``line-emission only'' and ``weights'' spectral images.}. This simulated image has sufficient spectral resolution, but fewer spectral channels, so the placeholder cube was created by concatenating the simulated cube four times, along the frequency dimension.

\begin{figure}
	\centering
	\resizebox{\columnwidth}{!}{\input{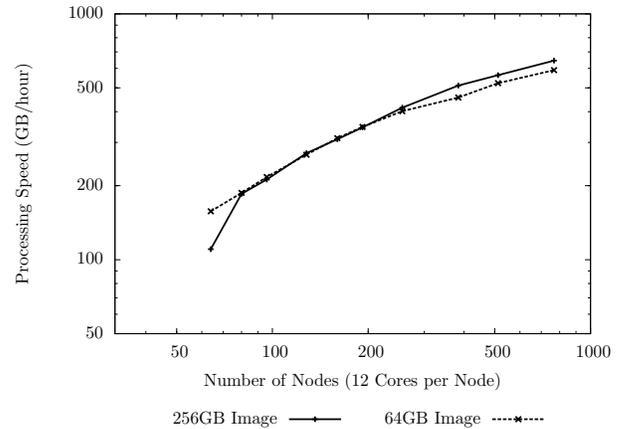}}
	\caption{The amount of data that the program can process per unit of time. This data is based on the use of 29,384 filter values in the convolution and either a 64GB or a 256GB file. The data size only includes the image data, not the weights data. This figure shows the rate at which the performance of the program scales with the number of cores used.}
	\label{processing_speed}
\end{figure}

Running the program with $768$ cores across $64$ nodes took $2$ hours $19$ minutes and $10$ seconds, consuming $1781$ core-hours. The system reported that the CPU utilisation was $99\%$. The rate at which the program can search an image and the manner in which it scales with the number of processors is shown in \figurename~\ref{processing_speed}. This figure can be used to estimate the processing time for images of different sizes and includes the processing speed as determined from the original 64GB simulated image, for comparison.

The processing time here is the mean time across ten runs of PGSF per node size. The exception is the file input time, which is treated separately from the other values, because it can vary greatly from one run of the program to another, depending on the amount of storage system bandwidth being consumed by other programs at that time. It is also possible for the image data to be stored in the cache of the storage system, which will cause subsequent runs to have an unusually fast input step. In practice, the program will experience congestion from other users but it is unlikely that the source finder will need to be run multiple times in succession, so it is unlikely for the image data to already be in the cache. For the purpose of demonstrating the scaling of the program with the number of cores used, the file input time was fixed to a value equal to the median time across different core counts that were not unusually small. Additionally, a small fraction of runs failed to complete, due to one or more of the nodes they were allocated either failed the \url{MPI_Init()} function, or they timed out and were terminated by the scheduling system. These are treated as outliers, and their data is not included here. These issues are considered further in the Discussion section.

\begin{figure}
	\centering
	\resizebox{\columnwidth}{!}{\input{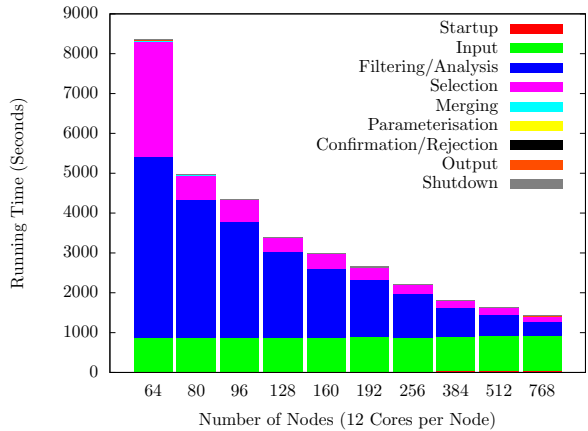}}
	\caption{The processing time as a function of the number of nodes and cores used. The minimum number of nodes is set by the amount of memory needed. The program needs enough nodes to ensure that there is sufficient memory to store the entire data set. The different colours denote different tasks in searching the image.}
	\label{running_time}
\end{figure}

The time required to search the images is broken down in \figurename~\ref{running_time}, with each section as detailed in the method section and the addition of two additional steps. These are the startup time and the shutdown time, which are included for completeness. These record the time taken to start and initialise the program, and the time taken to clean up and shutdown the program, respectively.

\section{DISCUSSION}
The suitability of a source finder for processing large HI spectral images is determined by a number of factors. The accuracy of the source finder is important as it determines the scientific usefulness of the results. The practicality of using the source finder to process large images is dependent on the computational performance of the source finder. As the size of the images increases the computational resources required to process them also increases, making the scalability of the source finding program a crucial aspect of its performance. The memory requirements of a program are also relevant, as they set the lower limit on the amount of resources required to process an image of a particular size. PGSF's performance in these measures is discussed below.

\subsection{Accuracy}
The accuracy of PGSF has been measured through the use of unit testing, and by testing the program using a simulated image. The accuracy with respect to finding sources is demonstrated in \figurename~\ref{accuracy}. For point sources the program has good accuracy for both completeness and reliability above a peak SNR of 5, although the reliability is greater than the completeness. Below a peak SNR of five the accuracy drops greatly. This is due to a large number of noise peaks being detected in the cube with a peak SNR of around 2-5. The completeness for sources with high SNR is less than $100\%$ because some true detections are being rejected due to their small spectral widths. The extended sources are detected with a slightly higher accuracy than the point sources, achieving good accuracy down to a peak SNR of around 4. The difference in accuracy between point and extended sources is likely because the point sources are small compared to the filters used, whilst the extended sources are a closer match to the filters used.

The completeness and reliability can be improved by using filter templates that better match the data being searched, using a more optimal set of parameters. The parameters used were selected to demonstrate the computational performance of the program. The accuracy results for a source finder are highly dependent on the parameters used, even a good detection algorithm can perform poorly with an inappropriate set of detection parameters. The values for the optimum parameters may also vary from one dataset to another. The extended sources are highly fragmented, particularly along the frequency axis, so a large merging width along the frequency axis was used. Whilst the merging radius along the frequency axis used was valid for the test data set being used, for real data a smaller radius is more appropriate, in order to prevent different objects from being merged into the same detection. The point sources are more compact, so these sources could be merged together correctly with a smaller merging radius.

\subsection{Computational Performance}
Testing the program shows that it can process a data cube in $24$ minutes using $9,216$ processing cores, although processing the data using $768$ cores uses less resources, at the cost of taking $2$ hours and $19$ minutes time. The survey time for WALLABY has yet to be finalised, but is expected to be eight to twelve hours per image. This means that the program can successfully search a WALLABY-sized image in significantly less time than the image takes to produce. Whether or not the entire image processing pipeline can create and analyse an image in less time than the observation used to produce the image will still depend on the performance of the other components in the pipeline, and the computational resources employed. It is still beneficial to improve the performance of the source finding program further as it can reduce the computing resources required to perform the survey or allow those resources to be used for other tasks in the image production pipeline.

In addition to the whole processing time \figurename~\ref{running_time} also identifies the components of PGSF that are the most time-consuming. Most of the steps consume a negligible amount of time. The filtering takes up the vast majority of the time, proportional to the number and size of the filters used, showing that this is the section that can benefit the most from optimisation, and that using additional nodes is effective in reducing the filtering time.

The time taken to read in the image from disk can vary greatly. The amount of time taken to read in the 256GB file varied from $9$ seconds to $46$ minutes, $39$ seconds. The difference in times is primarily caused by the extent to which the data is already in the system's cache from a previous run, and the extent to which other programs are accessing the storage nodes and using the network bandwidth. Across all runs that weren't outliers, PGSF took an average of $14$ minutes and $41$ seconds to read the file in, and this value is dependent on the bandwidth of the storage system not the number of cores used. Including outliers, the average time to read the file was $10$ minutes and $46$ seconds.

If PGSF were to be used as part of an image production pipeline, as opposed to a single program, then the file input time may be irrelevant if the image is already in memory from the previous steps of the pipeline. Depending on the manner in which the data is distributed after the previous step of the pipeline, the source finder may need to transpose or otherwise rearrange the data into a format that it can use. This rearrangement would itself take time, but it would not be limited by the storage system.

The other steps in the program can also have variations in processing time from one run to the next. Steps that involve communication can be slowed by other programs using the available bandwidth. Computation can vary in time due to differences in process scheduling on CPUs. The greatest differences in time from one run to the next, apart from the file input step, can be seen in the analysis step, because it is the longest. The convolution algorithm used has no communication, and minimal overhead, so the time taken should be inversely proportional to the number of cores used.

\subsection{Scaling}
It is important to analyse the manner in which the processing speed of PGSF varies with the number of cores used to process the data, as this is the most straightforward way of increasing the speed of the program. The data processing rate as a function of the number of processing cores is shown in \figurename~\ref{processing_speed}. It demonstrates that the speed generally increases as the number of cores increases, but the speedup is less than linear.

The single greatest consumption of processing time is the analysis step, which consists almost entirely of performing the convolution. This function is proportional to the number of voxels in the image multiplied by the total number of filter elements used. The other notable time-consuming steps in the program, reading the image into memory and calculating the mean and standard deviation, both have a time complexity that is proportional to the size of the image. Because all of the steps that take a considerable amount of time are proportional to the size of the image, the processing speeds shown in \figurename~\ref{processing_speed} can be used to estimate the time the program will take to search images of other sizes, when using the same  number and speed of processors and the same number of filter elements. This is demonstrated by also running the program on the original 64GB image, and observing that it achieves the same processing speed. The scaling of processing time against the image size will hold until the image is large enough that the computer system used to search it no longer has enough memory to store all the data structures needed by the program.

As the number of cores varies, it is possible for the main bottleneck of the program to occur in different components. In the sizes shown, the main bottleneck is the filter convolution algorithm. This algorithm has no communication between processes, the total amount of processing is constant with the number of processes, and its processing can be evenly divided between processes, so it will scale linearly with the number of cores used.

There is a potential for a bottleneck to form when reading the image in from storage. This step is limited by the bandwidth between the storage system and the compute nodes. If the data is stored in a distributed manner, and the network bandwidth is sufficient, then increasing the number of storage nodes would increase the bandwidth available to read the data and so increase the speed of the program. However, if the machine only has a limited number of storage nodes, the increasing the number of compute nodes used to process the image will not increase the rate at which the data is read and so the input time will remain constant. Additionally, the input step may take significantly longer if other programs on the supercomputer are using the storage system bandwidth at the same time as the source finder. The input step is the primary reason the processing speed increases less than linearly with the number of nodes.

The next potential bottleneck is in the local statistics function. This method requires that the halo data be exchanged once per filter used. The amount of data transfer grows larger in proportion to the total image size as the number of nodes increases. The computational work required stays constant as the number of nodes increases and is linearly parallelisable. As the number of processes used increases, the proportionally larger data transfer may come to dominate the computation. This per-filter data transfer could be avoided by having each node filter its own halo data, which would also involve increasing the size of the halo data to include both the range of the local statistics and the size of the filter. However, this implementation would significantly increase the amount of processing required, as there would be a constant amount of halo data per process that would need filtering, in addition to that node's portion of the image. The extra computation could cost more time than avoiding the data transfer saves, depending on the relative processing and network bandwidth of the machine being used. Another alternative is to use a global calculation for the mean and standard deviation. Although this method results in a different value for the threshold, it requires significantly less data transfer.

There is also a potential bottleneck in the functions that merge voxels into sources, and then parameterise them. The point at which the merging and parameterisation become significant contributers to the overall runtime of the program will vary with the settings used to run the source finder. The time spent filtering the data and performing statistics calculations depends on the size and number of the filters used and the input time is dependent on the size of the image data. In comparison, the time spent merging and parameterising sources depends on the number of detected voxels, which will in turn depend on the image data being searched, the parameters used to filter the image and particularly the detection thresholds used. Additionally, an increased number of detections require additional memory to store their information, which can potentially consume all the available memory on a machine and force data into swap space, significantly slowing the entire program.

In practice these functions are not a concern because they normally take a very small component of the overall time of the program. The program must detect an extremely large number of voxels for these segments to take a large portion of the processing time, in which case the number of objects found is so large that the vast majority of them are likely to be false detections. For example, the WALLABY survey is expected to find around $1,000$ detections per image. This figure is the result after the confirmation step, so a greater number of possible detections will pass through the voxel merging and parameterisation steps. Testing on Epic using the 256GB image resulted in PGSF finding $560,544$ voxels across $13,172$ sources post-confirmation, with the thresholding, voxel merging and parameterisation steps combined taking $0.4-2.3\%$ of the total running time, depending on the run. Thus PGSF should be capable of dealing with the number of sources expected from the WALLABY survey. In testing, this issue occurred primarily when using a particularly low detection threshold for the filters, but depending on the image data it is also possible for extended sources to have enough voxels that merging and parameterising them consumes a significant amount of processing time.

The scalability information of the program also suggests the optimal number of nodes to use when running PGSF. For example, increasing the number of nodes from $80$ to $512$ nodes, a times $6.4$ increase in computing resources, decreases the running time by a factor of $3$. If a cube needs to be searched quickly it would be reasonable to use a large number of nodes to reduce the processing time, at the cost of consuming a greater number of core-hours. If a longer delay is acceptable then it would be more efficient to use the minimum number of nodes per job. To search $1,000$ WALLABY images would take less than $139$ hours if the source finder were to use all $800$ nodes of the Epic supercomputer, with $10$ jobs executing in parallel on $80$ nodes each. In comparison, it would take approximately $397$ hours of processing using a single $768$ node job at a time.

The load is balanced by distributing the image evenly between the different processes. The most time-consuming tasks are reading in the data, convolving the data with the filters, and calculating the mean and standard deviation. The time required for a process to complete these tasks is proportional to the size of the image data assigned to that process, so the computational effort is evenly balanced between the nodes. Other functions are less evenly balanced. The time required by the thresholding algorithm increases with the number of voxels that are above the threshold. This value is data dependent, so it will not necessarily be balanced. More importantly, the time taken by the merging and parameterisation steps is proportional to the number of voxels and sources held by those nodes. Each node processes the sources that were found in its portion of the image, after merging sources that are present among multiple nodes. This means that the load balance of these steps is dependent on the distribution of potential sources across the image data. In practice, for the data tested these tasks take so little time that any load imbalance has little effect on the overall run time of the program. For a more complex parameterisation algorithm it may be beneficial to run the parameterisation, and consequently the confirmation task, in a separate program that better balances the work required between the nodes used.

\subsection{Memory}
The memory requirements for the program are approximately three times the size of the image being searched, or four times when using the weighted version of the algorithms. There are three data structures that consume almost all of the memory required by the program. The first two are the original copy of the image and the filtered copy of the image. Each one requires an amount of memory equal to the size of the image being searched, with a small amount of extra memory for the halo values. The third major use of memory is keeping track of the voxels and the number of different filters for which they are above the threshold. The actual amount of memory required depends on the data set, as memory is only required for voxels that are above the threshold for at least one filter.

Searching a larger image will require extra memory proportional to the increase in size of the image. This means that searching a larger image will require a proportionally larger number of nodes to supply the memory. The additional nodes would also provide additional computational power, but due to inefficiencies in scaling searching a larger image using a proportionally larger number of nodes would be slower than searching the original image.

A small fraction of runs for the 256GB image file failed to complete. This issue primarily affected the runs with small numbers of nodes, at 96 nodes and above no problems were encountered. Measuring the individual processes it appears that slowdowns occur in functions such as the filtering, where there is no communication and a uniform expected processing time. It is suspected that this is caused by some of the physical RAM in a node being unavailable to the source finder processes, which caused data for the source finder to be pushed into swap space. Jobs for the affected node sizes were also tested on Fornax, a smaller cluster with similar architecture but more memory per node. These issues were not encountered on that system.

This slowdown can be seen in \figurename~\ref{running_time} where the selection time is significantly higher for 64 nodes than for the other node sizes.  The extra time appears in the selection step because that is where the halo transfer occurs. During the halo transfer nodes must wait for their surrounding nodes for the data transfer to finish, so if one node is slow then its neighbours will wait for it in this step. In future work, it would be beneficial to add fault tolerance to the source finder to avoid or compensate for these problems.

\section{SUMMARY}
The Scalable Source Finding Framework detailed by this work provides an method for constructing parallel source finders, so that they can make use of HPC architecture. This was demonstrated by using SSoFF to write PGSF, which is capable of searching large images. Further algorithms can be implemented using SSoFF with limited concern for a distributed dataset, provided they operate on a local area of the image. There are some limitations the SSoFF and the presented source finder. The scalability of the framework is limited by the communication needs of the source finder, and the file input speed is limited by the bandwidth of the storage system independent of the processing nodes, which may significantly slow down the overall speed of searching an image. PGSF keeps memory overhead relatively low, at two to four times the size of the image, but with a large image file this can still impose a large memory requirement on the processing system. It is possible to reduce this memory requirement, but only at the cost of reading additional information from storage, slowing the program. Finally, other algorithms implemented using SSoFF will carry their own computational and communications costs, which will affect the processing speed of a source finder. Overall, SSoFF provides a suitable framework for writing source finders that make use of parallel HPC systems.

\subsection{Future Work}
There are a number of additions that could be made to SSoFF. Most notable is support for different file formats for the image data, such as FITS or HDF5. These are not yet included because the WALLABY survey has yet to finalise the file format they will use. With additional file formats the framework could also read in the metadata of an image, for use in parameterisation. Additionally, the time taken to read in file data and the manner in which it varies could be considered in greater detail in future work. The statistics functions could be expanded to include median-based statistics functions, for more robust calculations. The voxel merging code can be improved with the use of a more sophisticated method for choosing the voxels to merge into a single object.

PGSF can also be improved in a number ways, indeed as a framework SSoFF is designed to ease the addition of functionality to a source finder. Different analysis techniques could be used to select voxels. The performance of the convolution algorithm shown could be improved by porting it to an accelerator device, such as a Graphics Processing Unit (GPU). The parameterisation algorithm could be improved, particularly with the use of metadata, and the confirmation step could use more complex criteria for its decision making.

\begin{acknowledgements}
The authors thank Russell Jurek for his assistance with parameterisation and Tobias Westmeier for his assistance in preparing this paper. This work was supported by iVEC through the use of advanced computing resources located at iVEC@Murdoch and iVEC@UWA.
\end{acknowledgements}


\begin{thebibliography}{12}
\expandafter\ifx\csname natexlab\endcsname\relax\def\natexlab#1{#1}\fi

\bibitem[{{Fl{\"o}er} \& {Winkel}(2012)}]{2d_1d_wavelet_source_finder}
{Fl{\"o}er}, L., \& {Winkel}, B. 2012, Publications of the Astronomical Society
  of Australia, 29, 244

\bibitem[{{Jurek}(2012)}]{cnhi_source_finder}
{Jurek}, R. 2012, Publications of the Astronomical Society of Australia, 29,
  251

\bibitem[{Koribalski \& Staveley-Smith(2009)}]{WallabyProposal}
Koribalski, B.~S., \& Staveley-Smith, L. 2009, Proposal for WALLABY: Widefield
  ASKAP L-band Legacy All-sky Blind surveY, Available from
  \url{http://www.atnf.csiro.au/research/WALLABY/proposal.html}, Last Visited 9
  September 2011

\bibitem[{Meyer {et~al.}(2004)Meyer, Zwaan, Webster, Staveley-Smith,
  Ryan-Weber, Drinkwater, Barnes, Howlett, Kilborn, Stevens,
  {et~al.}}]{HIPASSCatalogue}
Meyer, M., {et~al.} 2004, \mnras, 350, 1195

\bibitem[{{OpenMP Architecture Review Board}(2008)}]{openmp08}
{OpenMP Architecture Review Board}. 2008, {OpenMP} Application Program
  Interface Version 3.0

\bibitem[{{Popping} {et~al.}(2012){Popping}, {Jurek}, {Westmeier}, {Serra},
  {Floer}, {Meyer}, \& {Koribalski}}]{source_finder_comparison}
{Popping}, A., {Jurek}, R., {Westmeier}, T., {Serra}, P., {Floer}, L., {Meyer},
  M., \& {Koribalski}, B. 2012, Publications of the Astronomical Society of
  Australia, 29, 318

\bibitem[{Serra {et~al.}(2012)Serra, Oosterloo, Morganti, Alatalo, Blitz, Bois,
  Bournaud, Bureau, Cappellari, Crocker, Davies, Davis, de~Zeeuw, Duc,
  Emsellem, Khochfar, Krajnović, Kuntschner, Lablanche, McDermid, Naab, Sarzi,
  Scott, Trager, Weijmans, \& Young}]{s+c_source_finder}
Serra, P., {et~al.} 2012, Monthly Notices of the Royal Astronomical Society,
  422, 1835

\bibitem[{{The MPI Forum}(1993)}]{mpi}
{The MPI Forum}. 1993, in Proceedings of the 1993 ACM/IEEE conference on
  Supercomputing, Supercomputing '93 (New York, NY, USA: ACM), 878--883

\bibitem[{{Westerlund} {et~al.}(2012){Westerlund}, {Harris}, \&
  {Westmeier}}]{sfae}
{Westerlund}, S., {Harris}, C., \& {Westmeier}, T. 2012, Publications of the
  Astronomical Society of Australia, 29, 301

\bibitem[{Whiting \& Humphreys(2012)}]{askap_source_finding}
Whiting, M., \& Humphreys, B. 2012, Publications of the Astronomical Society of
  Australia, 29, 371

\bibitem[{Whiting(2012)}]{duchamp_paper}
Whiting, M.~T. 2012, Monthly Notices of the Royal Astronomical Society, 421,
  3242

\bibitem[{Winkel(2008)}]{WinkelThesis}
Winkel, B. 2008, PhD thesis, Universit{\"a}t Bonn

\end{thebibliography}

\end{document}